\documentclass[traditabstract]{aa} 
\usepackage{graphicx}
\usepackage{txfonts}
\usepackage{natbib}
\usepackage{deluxetable}

\newcommand{\vsini}{\mbox{$v \sin i$}}
\newcommand{\kms}{\mbox{km\, s$^{-1}$}}

\begin{document}


\title{MML~53: a new low-mass, pre-main sequence eclipsing binary in the Upper Centarus-Lupus Region discovered by SuperWASP}
\titlerunning{MML~53: A new PMS EB discovered by SuperWASP}

\author{
L.~Hebb\inst{1,2}\fnmsep\thanks{email: \mbox{leslie.hebb@vanderbilt.edu}}
\and
H.C.~Stempels\inst{3,1}
\and
S.~Aigrain\inst{4}
\and
A.~Collier-Cameron\inst{1}
\and
S.T.~Hodgkin\inst{5}
\and
J.M.~Irwin\inst{6}
\and
P.F.L.~Maxted\inst{7}
\and
D.~Pollacco\inst{8}
\and
R.A.~Street\inst{9}
\and
D.M.~Wilson\inst{7,10}
\and
K.G.~Stassun\inst{2,11}
}

\institute{
School of Physics and Astronomy, University of St Andrews, North Haugh, St Andrews KY16 9SS, UK 
\and
Department of Physics \& Astronomy, Vanderbilt University, Nashville, TN 37235, USA  
\and
Department of Physics \& Astronomy, Box 516, SE-751 20 Uppsala, Sweden 
\and
Department of Physics, University of Oxford, Keble Road, Oxford OX1 3RH, UK 
\and
Institute of Astronomy, Cambridge University, Madingley Road,  Cambridge CB3 0HA, UK 
\and
Harvard-Smithsonian Center for Astrophysics, 60 Garden Street, Cambridge, MA 02138 USA 
\and
Astrophysics Group, Keele University, Staffordshire ST5 5BG, UK
\and
Astrophysics Research Centre, School of Mathematics \&\ Physics, Queen's University, University Road, Belfast BT7 1NN, UK 
\and
Las Cumbres Observatory, 6740 Cortona Dr. Suite 102, Santa Barbara, CA 93117, USA  
\and
Center for Astrophysics \& Planetary Science, University of Kent, Canterbury CT2 7NH, UK
\and
Department of Physics, Fisk University, 1000 N. 17th Ave., Nashville, TN 37208, USA 
}

\date{Received; accepted }

\abstract
{We announce the discovery of a new low-mass, pre-main sequence eclipsing binary,
MML~53.  Previous observations of MML~53 found it to be a
pre-main sequence spectroscopic multiple associated with the 15-22~Myr Upper Centaurus
Lupus cluster.
We identify the object as an eclipsing binary for the first time
through the analysis of multiple seasons of time series photometry from
the SuperWASP transiting planet survey. Re-analysis of a single archive spectrum shows
MML~53 to be a spatially unresolved triple system of young stars which all exhibit
significant lithium absorption.
Two of the components comprise an eclipsing binary
with period, $P = 2.097891(6) \pm 0.000005$ and mass
ratio, $q\sim 0.8$.  Here, we present the analysis of the discovery data. }


\maketitle


\section{Introduction}

Eclipsing binaries are systems
in which the masses, radii, temperatures, and absolute luminosities can be
measured for two stars with the same age and metallicity \citep{andersen91,youngarnett01}.  
In addition, apsidal motion measurements, if available, can further constrain 
interior stellar physics \citep{youngarnett05,schwarz57}.
Therefore, these objects are effective tools used extensively 
for calibrating and testing theoretical stellar evolution models.  

Pre-main sequence eclipsing binaries are particularly important.
Young stars are changing rapidly as they contract onto the main sequence and heat up, 
so an individual solar-type star covers a large range of temperatures and radii 
in its first 50~Myr.  Furthermore, young stars often have star spots, coronal
activity, and/or circumstellar dust which complicates the physics of their formation
and early evolution.  Thus, empirical measurements of fundamental properties 
of young stars with a range of masses and ages are needed to fully constrain
the pre-main sequence evolution and understand how the properties of young stars 
are affected by other parameters (e.g.~magnetic fields, dust, metallicity, planets, multiplicity).

Pre-main sequence eclipsing binaries are the only objects which can provide 
precise measurements of the most fundamental physical properties of young stars,
however these objects are extremely rare.  There are only 6 known
low-mass pre-main sequence eclipsing binaries with $M<1.5$~M$_{\odot}$:
RXJ~0529.4+0041A \citep{covino00,covino04},
V1174~Ori \citep{stassun04}, 2MASS~J05352184-20130546085
\citep{stassun06,stassun07}, JW~380 \citep{irwin07}, Par~1802 \citep{cargile,stassun08},
and ASAS~J052821+0338.5 \citep{stempels08}. In
addition, EK~Cep \citep{popper87}, TY~CrA\citep{casey98}, and RS~Cha \citep{rscha1,rscha2} are known 
to be higher mass eclipsing systems with at least one PMS component.

In this paper, we announce the discovery of a new low-mass PMS eclipsing binary, and
the first such object discovered outside of the Orion star forming region.
We describe the observations that were used to identify the object
as a PMS eclipsing binary (\S\ref{sec:observations}).  Using the discovery data, we determine
initial orbital parameters for the system and approximate physical properties of its component 
stars (\S\ref{sec:results}).  Finally, we discuss future 
observations needed to fully analyse this important system (\S\ref{sec:summary}).

\section{Observations}
\label{sec:observations}

TYC 7310-503-1 ($\alpha=$14:58:37.7, $\delta=$-35:40:30.4), MML~53, is
a late-type star spatially located in the region of the 
Scorpius Centarus OB association complex.  It was identifed
as an X-ray source by ROSAT (RXJ1458.6-3541) which lead
to further spectroscopic study by several groups with the aim of
identifying young stars and mapping the overall star forming
region \citep{Wichmann,Wichmannb,Mamajek,Torres,White}.  The object
was first identified as a T~Tauri star (K3-type) through the 
detection of Li~{\sc I}~$\lambda6708$ absorption by
\citet{Wichmannb}.  
H$\alpha$ emission and significant lithium absorption 
were detected in all  subsequent spectroscopic observations,
therefore MML~53 is confirmed to be a young, pre-main sequence (PMS) object.

\citet{Mamajek} also kinematically and spatially defined
MML~53 as a member of the 15-22 Myr old Upper Centaurus Lupus (UCL)
sub-association with a 93\% probability.  The authors classified
the object as a K2IVe star based on several spectroscopic 
indices and determined a kinematic parallax of 
$7.36\pm0.77$~mas ($136^{+16}_{-13}$~pc).  These values and the 
high probability of membership in UCL are confirmed in more recent work
by these authors (E.~Mamajek priv communication).  \citet{Torres} made a
rotation measurement of $\vsini = 30.8\pm 3.1$~\kms and were
the first to identify the target as a possible spectroscopic triple system
with a systemic radial velocity, $\gamma=2.0$~\kms.
\citet{White} detected a double-lined nature for the object
with a single observation and measured the equivalent
width of lithium in the individual primary and secondary components
(122 and 222 m\AA). \citet{White} do not
explicitely state which binary component is associated with the
reported Li I EW values since they are unable to derive an orbit from their 
spectrum.  The approximate values
we derive for Li EW in Sect.~\ref{sec:modelspec} 
are consistent with these measurements if we assume 
the primary star has the larger EW (222 m\AA).

As a known PMS object, MML~53 has also been included in the Spitzer
Legacy project investigating the formation and evolution of planetary
systems (FEPS) \citet{Meyer}.  Spitzer observations of the object were obtained
in all IRAC and MIPS bands from 3.6-70~$\mu$m \citep{Carpentera},
but no primordial or a debris disk was detected \citep{Silverstone,Carpenterb}.
In Table~\ref{tab:litparms}, we list the known properties of the star
obtained from the literature.

\begin{table}
\caption{Properties of MML 53 obtained from the literature. }
\begin{center}
\begin{tabular}{ccc}
Parameter    & Value & Reference \\
\hline\\
${\rm RA (J2000)}$    &  14:58:37.70             &      \\
${\rm Dec (J2000)}$   & -35:40:30.4              &       \\
${\rm VT}$            &  $10.88$ mag           & 1        \\
${\rm I}$             &  $9.85\pm 0.04$ mag    & 2          \\
${\rm J}$             &  $8.639\pm 0.024$ mag  & 3          \\
${\rm H}$             &  $8.062\pm 0.055$ mag  & 3         \\
${\rm K}$             &  $7.870\pm 0.026$ mag  & 3         \\
                      &                                &  \\
${\mu_{\rm RA}}$      & $-22.3 \pm 1.1$ mas yr$^{-1}$ & 4  \\
${\mu_{\rm DEC}}$     & $-25.0 \pm 1.1$ mas yr$^{-1}$ & 4  \\
                      &                                &  \\
${\rm SpT}$           & {\rm K2IVe}                & 5     \\
${\rm \pi}$           & $7.36\pm 0.77$ mas         & 5     \\
${\rm \gamma}$          & $2.0 \pm 3.1$ \kms         & 6     \\
\vsini              & $30.8 \pm 3.1$ \kms        & 6     \\
\hline
\end{tabular}
\end{center}

\begin{list}{}{}
\item[(1)] TYCHO \citep{TYCHO}
\item[(2)] DENIS \citep{DENIS}
\item[(3)] 2MASS \citep{2MASS}
\item[(4)] NOMAD \citep{NOMAD}
\item[(5)] \citep{Mamajek}
\item[(6)] SACY \citep{Torres}
\end{list}

\label{tab:litparms}

\end{table}

\subsection{SuperWASP Photometry}

In addition to the targeted surveys of young stars mentioned above, 
MML~53 was observed in the field-of-view of 
the SuperWASP transiting planet survey \citep{swasp_instr}.  SuperWASP is a wide-field photometric
variability survey designed to detect transiting gas-giant planets around bright
main sequence stars.
The survey cameras repeatedly ($\sim 8$~minute sampling) observe bright stars ($V\sim 9-13$) 
at high precision (1\%) using a broad V+R-band filter. 

MML~53 was observed with the WASP-South telescope and instrumentation \citep{swasp_instr,wasp4}
in the spring observing season from 2006 to 2008.  3492 photometric
data points were obtained between 4 May - 30 July 2006; 5602
measurements were made between 16 Feb 2007 - 19 July 2007;
and 2886 observations were taken from 18 Feb 2008 - 17 Apr 2008.  
All data sets were processed independently with the standard SuperWASP data reduction
and photometry pipeline \citep{hunter}.  The light curves, which were detrended along with 
other stars in the field using the SysRem algorithm \citep{tamuz},
were then run through our implementation of the box least squares (BLS) 
algorithm \citep{kovacs,hunter}.  The BLS algorithm is designed 
specifically to detect square shaped dips in brightness in an otherwise flat light curve. 
It is very effective at detecting transits by extrasolar planets \citep{tingley} and eclipsing binary 
systems \citep{hartman}.    

\begin{figure}
  \centering
  \includegraphics[angle=0,width=\columnwidth]{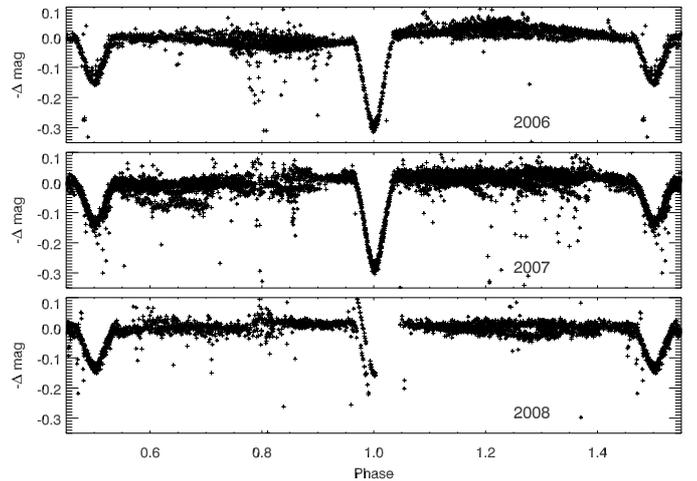}
  \caption{WASP-South photometry of MML~53 from 2006-2008 obtained in a
   broad V+R-band filter.  The data 
   are phase-folded with the ephemeris derived from the eclipse
   model, $HJD = 2454301.3252(7) \pm 0.002 + 2.097891(6) \pm 0.000005 $.
  }
  \label{fig:swasplc}
\end{figure}

Using the BLS algorithm, we identified MML~53 as having an eclipsing binary light curve
with a period of $\sim 2.1$~days.
Figure~\ref{fig:swasplc} shows the phase-folded SuperWASP data.
The target is isolated, and the field-of-view 
shows no other nearby stars that could have contaminated the large photometry 
aperture ($\sim 48^{\prime\prime}$) causing the resulting light curve.

\begin{figure}
 \centering
 \includegraphics[angle=0,width=\columnwidth]{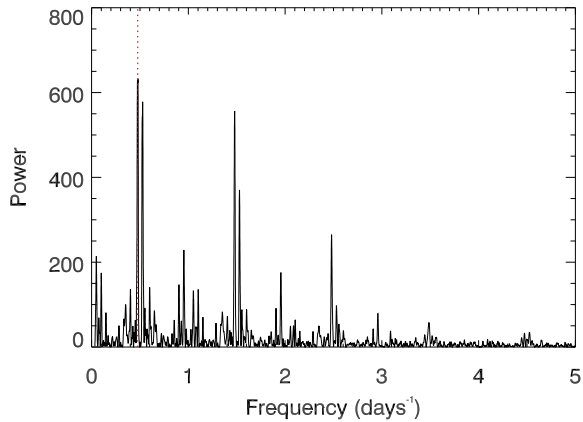}
 \caption{Lomb-Scargle periodogram results from the 2006 SuperWASP 
  photometry with the in-eclipse data removed.  Periods between
  0.2-30~days were searched.  A periodic signal was
  detected at $P\sim 2.09$~days in all 3 seasons of data, but
  the signal is strongest in the 2006 data shown here.  The dotted line
  denotes the orbital period of the binary, $P=2.097891(6) \pm 0.000005$.
  }
 \label{fig:pdgram}
\end{figure}

The primary and secondary eclipses are apparent in the light curve as well as
sinusoidal out-of-eclipse variability which is particularly evident in the 2006 data.
Brightness variations due to photospheric star spots are often present on 
young and/or active stars and can be used to measure stellar rotation periods.
Thus, we investigated the presence of rotational variability on MML~53 to
determine if the system is tidally synchronized.
We measured the period of the out-of-eclipse variability by applying the Lomb-Scargle
algorithm \citep{scargle,horne} to each season's light curve independently after removing all the
in-eclipse data.  We searched periods between 0.2-30~days.  
In the 2006 data, there is a highly signficant
sinusoidal signal with a period, $P=2.09$~days (see Fig.~\ref{fig:pdgram}).  
Aliased peaks due to the window function are also present at a reduced amplitude.  
In the 2007 and 2008 data sets, the signal is weaker, but still present.   
This period is matched to the orbital period of the binary suggesting the
out-of-eclipse variability is due to starspots on the surface of one or both
of the binary components and that the component stars are rotating synchronously with the orbit.   
The sinusoidal variability is changing phase and amplitude on the timescale of 
months which manifests as visible scatter in the light curve 
(see Fig.~\ref{fig:swasplc}) indicating a slow drift in the starspot pattern, 
but the measured rotation period is consistent in all the seasons of data.
For a short period binary like MML~53, we expect the components to be rotating 
synchronously despite their young age \citep[][and references therein]{zahn,mathieu}.

\subsection{Archive FEROS Spectrum}
\label{sec:feros}

After determining the system was an eclipsing binary, we searched for existing
archival spectroscopic data on the object and found one high resolution
($R\sim$~50 000) spectrum located in the European Southern Observatory
(ESO) archive. The spectrum was obtained
at heliocentric Julian date, $HJD=2453909.622300$ with the FEROS spectrograph on
the 2.2m ESO telescope at La Silla (observing program 077.C-0138(A), which forms
the basis of the SACY survey, see \citet{Torres}). 
We reduced this spectrum with the echelle data reduction package {\sc REDUCE} \citep{piskunov02}, 
using calibration data obtained on the same night. 
We also downloaded and reduced the spectrum of a radial velocity standard, HD 10700. 
Our final reduced spectrum covers the wavelength range of 3765--8862\AA\ and has
a signal-to-noise of $\sim 30$ per pixel at 6000\AA.

\begin{figure}
  \centering
  \includegraphics[angle=90,width=\columnwidth]{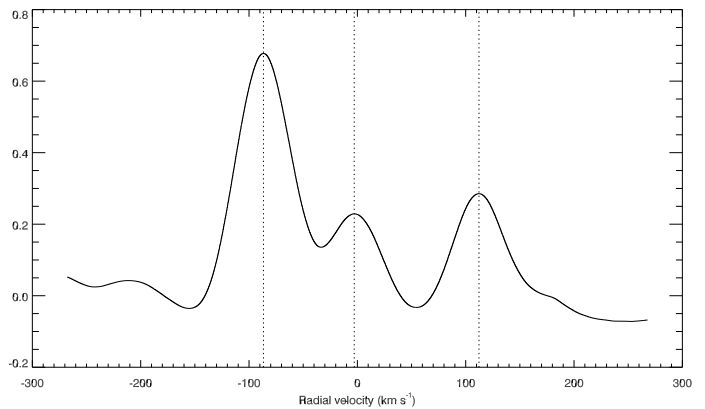}
  \caption{Cross correlation function of the FEROS spectrum of
   MML 53 obtained on 2006 June 23 compared to the template.
   The analysis  
   shows three different spatially unresolved components in the system
   with radial velocities of $-85.8$, $111.1$ and $-3.5$~\kms
   for the primary, secondary, and tertiary components, respectively.
  }
  \label{fig:ccf}
\end{figure}

When the spectrum was examined by eye, multiple stellar components were clearly
present as previously reported \citep{Torres,White} and as expected for an
eclipsing binary where both the primary and secondary eclipses are visible. 
However, \citet{White} report the system as a double-lined spectroscopic binary
(SB2) while \citet{Torres} suggest the possibility of a third component (SB3). 
To determine whether the system contains a third star and to measure the
radial velocities of the components, we perform a standard cross-correlation (Fig.~\ref{fig:ccf}) 
and also apply our implementation of the
least-squares deconvolution (LSD) algorithm to the MML 53 spectrum.
Least-squares deconvolution allows for deriving a very high signal-to-noise
average absorption line profile of the system by properly combining the many
individual lines throughout the entire spectrum \citep{donati_lsd}.  This technique can be used 
in spectroscopic binary work to identify faint unresolved companions. 
Both the cross-correlation function (CCF) and the result of the LSD analysis
clearly show three stellar components present in the spectrum.
We measure their radial velocities
to be $-85.8$, $111.1$ and $-3.5$~\kms, for the primary, 
secondary and tertiary, respectively.

\section{Analysis}
\label{sec:results}

\subsection{Light Curve Modelling}
\label{sec:lcparams}

In order to derive an ephemeris for the system 
and initial parameters for the individual eclipsing binary components,
we first generated a single recitified light curve from the discovery data
by removing the out-of-eclipse variability on each night and
combining all 3 seasons of SuperWASP photometry.
We fit a second order polynomial to the out-of-eclipse photometry on each 
night and applied the polynomial fit to all the data obtained that night.
We excluded from the rectified light curve any data taken on 
nights in which there was no out-of-eclipse photometry.
The fitting functions are not physical, and we made no attempt 
to model the starspot variability because the existing data are
only in a single band and therefore any physical model would be too full of
degeneracies to provide a useful result.  The data include 17 near complete
primary or secondary eclipses which show the eclipse minimum and 
out-of-eclipse photometry before and/or after the eclipse.
The final rectified light curve contains 11808 photometric measurements.

We fit the rectified light curve using the JKT Eclipsing Binary Orbit Program (EBOP)
\citep{popperetzel,southworth}.  The program determines the optimal
model light curve that matches the observed photometry and reports the binary
parameters for the model.  The algorithm on which it is based is only valid when analysing
well-detached eclipsing binaries in which the tidal distortion is small (i.e. nearly spherical
stars with oblateness $< 0.04$).  This is the case for MML~53.
The derived light curve parameters include the period, $P$, time of minimum light, $T_0$, 
surface brightness ratio in the SuperWASP filter, $J_{V+R}$, relative sum of the radii, $(R_1+R_2)/a$, 
inclination angle, $i$, eccentricity, $e$, and angle of periastron, $\omega$.
The routine takes into account the effects of limb darkening, gravity brightening, and
reflection effects and can account for the presence of light from a third component.  

We adopted the quadratic limb darkening coefficients from \citet{claret} 
using the temperatures for the eclipsing components that are defined below
in Sect.~\ref{sec:modelspec}, $T_{\rm eff,1}=4886$~K and $T_{\rm eff,2}=4309$~K.
We initially ran the fitting program allowing the eccentricity
to be a free parameter, however the resulting value was within 3$\sigma$ of zero, 
and there is no further evidence in the light curve for a non-circular orbit.

\begin{figure}
  \centering
  \includegraphics[angle=0,width=\columnwidth]{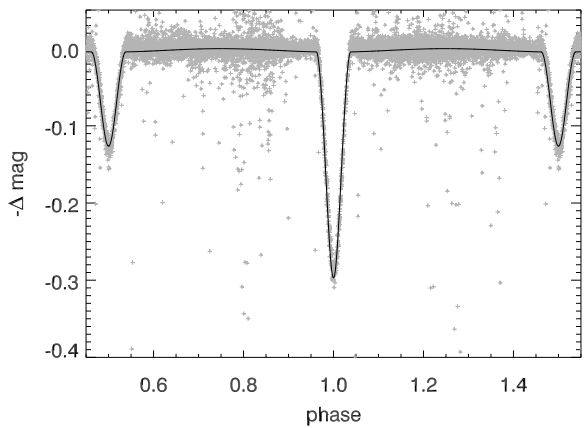}
  \caption{Rectified SuperWASP light curve phase-folded with ephemeris, 
  $HJD = 2454301.3252(7) \pm 0.002 + 2.097891(6) \pm 0.000005 $. 
  The EBOP model fit is overplotted (solid black line).}
  \label{fig:lcrect}
\end{figure}

Thus, in the final run of the light curve modelling program, we fixed the eccentricity and
angle of periastron to zero.  The result is shown in Fig.~\ref{fig:lcrect} with
the best fitting model overplotted on the final rectified light curve.
The fit provides a precise system ephemeris of:   
$${\rm Min (HJD)} = 2454301.3252(7) \pm 0.002 + 2.097891(6) \pm 0.000005 E $$

As an additional check, we determined
the ephemeris from each individual season of data, and found
the periods agree to within $5\times10^{-6}$~days.  However, the epoch of mininum light 
varies by $\sim 0.001$ in phase ($\sim 3$~min) between the 2006 and 2007-08 seasons.
This offset could be caused by effects from the third component (e.g. direct
gravitational influence, light travel time variations), if 
ultimately confirmed with current epoch eclipse data.  Based on the variations
from season to season, we adopt an uncertainty
on the time of minimum light of $0.002$~days and an uncertainty on 
the period of $0.000005$~days.

The model fit also gives a measurement for the relative sum of the radii, $(R_1+R_2)/a = 0.260$, 
the inclination angle, $i = 83.1^{\circ}$, and the surface brightness ratio in the 
SuperWASP filter, $J_{V+R} = 0.461$.  These values depend on the flux 
contribution of the third component which we estimate to be 15\% of the total 
luminosity based on our self consistent analysis of the light curve and the FEROS spectrum
described in Sect.~\ref{sec:modelspec}.  However, we stress that the fitted parameters will  
change when definitive temperatures for all three components and accurate luminosity ratios
can be derived through spectral disentangling of multiple high resolution, high 
signal-to-noise spectra of MML~53.  Furthermore, star spots which we know are 
present are likely to have an affect on these parameters, and multi-band photometry is
necessary to derive a comprehensive solution that models both the spots
and the eclipses.

\subsection{Initial Estimates of the Masses and Radii of the MML~53 Eclipsing Components}
\label{sec:components}

Using the radial velocity measurements of the primary and secondary star from the single archive spectrum, the
precise ephemeris and inclination angle estimate from the photometry, and
the systemtic radial velocity from the literature,
we derive approximate masses for the individual eclipsing components
of MML~53.  The FEROS observation was obtained very near to quadrature
at an orbital phase of $0.287$.  With the phase of the observation fixed, the 
radial velocity measurements for the primary and secondary (-85.8, +111.1~\kms) 
constrain the amplitudes of the sinusoidal (circular orbit) 
radial velocity curves for the two components, $K_1$ and $K_2$.  
Adopting 2.0~\kms \citep{Torres} for the systemic RV, we
find $K_1 = 90$ and $K_2=112$~\kms (assuming no uncertainty in the RV
measurements).  This gives a mass ratio for the system, $M_2$/$M_1=0.8$.  Taking
the period and inclination angle derived from the SuperWASP eclipse 
photometry, we find individual masses of $M_1\sim 1.0$~M$_{\odot}$, $M_2\sim 0.8$~M$_{\odot}$, and an 
orbital separation, $a\sim 8.4$~R$_{\odot}$.  

The derived mass estimates are consistent with the pre-main sequence status
of MML~53 determined from the observed youth indicators previously discussed.
Young stars should be comparatively large in radius, since they are still contracting onto the main sequence.
For MML~53, the sum of the radii measured from the light curve ($R_1+R_2 = 2.2$~R$_{\odot}$) is 
$\sim 30$\% larger than what is
expected for two main sequence stars (age of 300~Myr) with masses of 1.0 and 0.8~M$_{\odot}$ 
according to theoretical stellar evolution models \citep{baraffe98}.  

Finally, we note the properties of MML~53 derived here are only approximations, and a complete 
radial velocity curve will be
required to determine accurate masses for the stellar components of MML~53.  
Furthermore, the third component, which cannot be studied in detail with the existing data  
could also affect the radial velocity measurements if it is found to be gravitationally
bound to the system, which is still unknown.

\subsection{Composite Model Spectrum: Temperatures and Lithium Abundances}
\label{sec:modelspec}

Despite the lack of time-series spectra, we perform a joint analysis of the 
single observed FEROS spectrum and
the SuperWASP light curve in order to approximate the effective temperatures, 
radii and relative luminosities of the three stellar components of MML~53.
To do this, we derive a three-component model spectrum (based on synthetic \citet{kurucz} model atmospheres)
that is a reasonable fit to the observed spectrum and is also consistent
with the constraints from the light curve.  The model for each
component is defined by its effective temperature
and its luminosity relative to the total system luminosity at $\sim 6700\AA$ (R-band).
These properties are interdependent, so we iterate until they provide
a self consistent solution.

\begin{figure}
  \centering
  \includegraphics[angle=90,width=\columnwidth]{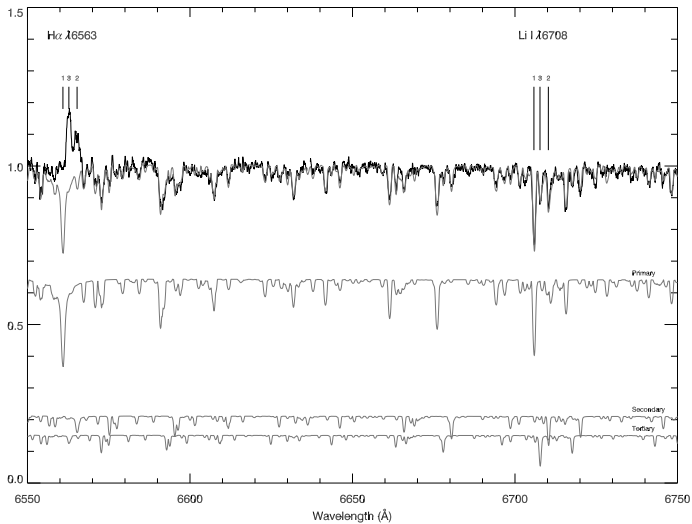}
  \caption{FEROS spectrum of the order containing H$\alpha$ and the Li~{\sc i} doublet
  at 6708\AA.  The solid black line is the observed spectrum.  Overplotted
  is the three-component model spectrum (grey line).  The model is a good
  match to the observed spectrum except near H$\alpha$ where 
  MML~53 shows emission and Li~{\sc i} where it shows significant absorption.
  Plotted below are the three individual synthetic spectra with temperatures 
  of 4886~K, 4309~K, and 4130~K for the primary, secondary and tertiary.  
  They are scaled according to the derived R-band luminosity ratios of 0.64:0.21:0.15
  and shifted by the velocity offsets found in the CCF analysis. 
  }
  \label{fig:specfig}
\end{figure}

After several iterations, we find temperatures of 4886~K, 4309~K, and 4130~K for
the primary, secondary and tertiary, respectively.  In addition, the fractional
R-band luminosities we use to scale the model spectra are 0.64:0.21:0.15 for the
three components.  These values provide a self consistent solution 
in which the third light input value for the light curve model 
is reproduced by the final temperature and radius of the tertiary.
In addition, the properties of the eclipsing components are consistent
with the light curve model.  Finally, these values
result in a composite model spectrum which is a good match to the 
absorption line depths of all three components in the observed spectrum
in the region between H$\alpha$ and Li~{\sc i}.
In Fig.~\ref{fig:specfig}, we show this region of the observed spectrum 
(black solid line) overplotted with the composite model defined by our best fitting
stellar properties.  
The model assumes a solar metallicity for all three components, and we adopt
rotational velocities for the stars of $\vsini = 29.0$, $23.2$, and $20.0$~\kms.  
Note that the $\vsini$ values for the primary and the secondary are determined
from the assumption of corotation and are reasonably consistent with the value
of 30.8~\kms measured by \citet{Torres}.

As can be seen in Fig.~\ref{fig:specfig}, MML~53 shows significant 
Li~{\sc i} absorption in all three components and H$\alpha$ in emission for the 
secondary and tertiary.  
We measure the equivalent width (EW) for each Li line in the combined spectrum 
to be 236, 75, and 82$\pm$~10~m\AA\ for the primary, secondary and tertiary, respectively.  
We then correct each of these values for the increased
contiuum due to the other components to derive corrected equivalent widths of EW(Li)$=369\pm 15,\ 356\pm 47, 550\pm 67$~m\AA\ 
which we translate into lithium abundances of ${\rm log}\ n\ ({\rm Li}) = 3.2$,\ $2.3$,\ and $3.0$.

The uncertainties on the lithium EWs quoted above do not include the systematic
error in the relative luminosity ratios which are difficult to estimate. 
The lithium abundance of the tertiary is unexpectedly larger than that of the secondary
which suggests some inaccuracies in our derived temperatures and luminosities.  
However, with additional observations of time series spectra, we will be
able to apply spectral disentangling which will allow for deriving accurate 
temperatures for the individual components and
measuring Li equivalent widths to greater precision.  

\section{Summary}
\label{sec:summary}

We report on the discovery of a new pre-main sequence ecipsing binary, MML~53,
located in the Lupus star forming region.  There are very few such objects known,
but they are extremely important for informing and calibrating stellar
evolution models at young ages where stars are changing rapidly
as they contract onto the main sequence.  This is only the seventh low-mass PMS 
eclipsing binary to be discovered.

Previous studies found the object to be a young spectroscopic multiple system
associated with the $15-22$~Myr Upper Centaurus Lupus cluster.
Through our analysis of SuperWASP photomery and a single archive spectrum, 
we determine MML~53 to be a triple system of young stars where two of the components
are eclipsing.  All three components show significant lithium absorption, and H$\alpha$
emission is seen in the secondary and tertiary.

We determine the orbital period of the eclipsing pair
to be $P=2.097891(6) \pm 0.000005$ using many primary and secondary eclipses observed
with the SuperWASP instrument over several years.  The light curve also shows out-of-eclipse 
variability consistent with star spots on the surface of at 
least one of the eclipsing components which is rotating synchronously with the orbit.

Analysis of the existing data suggests the eclipsing component stars have masses 
of $M_1\sim 1.0$~M$_{\odot}$ and $M_2\sim 0.8$~M$_{\odot}$.  Therefore, MML~53 is
a slightly older analogue of the V1174 Ori PMS eclipsing binary found in 
the $\sim 10$~Myr Orion OB1c sub association ($M_1=1.0$~M$_{\odot}$ and $M_2=0.73$~M$_{\odot}$).
When MML~53 is fully analysed, relative comparison of the two objects will allow for 
constraining the PMS evolution of a solar-type star to very high accuracy.  In addition,
because MML~53 has an expected age older than any of the other known low-mass PMS
eclipsing binaries, it will be valuable for exploring a slightly later stage of PMS evolution.

However, additional data is required to derive precise fundamental properties for the 
components of MML~53 before comparing to stellar evolution models and other
PMS eclipsing binaries.  Time-series spectroscopic data are needed to define a 
complete radial velocity curve which will allow for determining masses of the  
primary and secondary components.  
Additional high resolution, high signal-to-noise spectra will allow for determining
precise temperatures and relative luminosities of the three components stars.  
Multi-band photometry will ultimately provide a complete solution for the
system with individual masses, radii, temperatures and luminosities for
two young, pre-main sequence stars.

\begin{acknowledgements}
The SuperWASP Consortium consists of astronomers primarily from the Queen's University Belfast,
St Andrews, Keele, Leicester, The Open University, Isaac Newton Group La Palma and
Instituto de  Astrof{\'i}sica de Canarias. The SuperWASP Cameras were constructed
and operated with funds made available from Consortium Universities and the UK's Science and
Technology Facilities Council. 

K.G.S. acknowledges support from NSF Career award AST-0349075 as well as a Cottrell Scholar award from the Research Corporation. We also acknowledge the support of the Vanderbilt International Office for fostering the collaboration between Queens University Belfast and Vanderbilt University.

The results presented here are based on ESO observations obtained as part of programme ID 077.C-0138.   
The data were taken with ESO Telescopes at the La Silla Observatory and were obtained from 
the ESO archive.

L.H.H. would like to thank Eric Mamajek for helpful information about the membership of this object.

\end{acknowledgements}

\end{document}